\documentclass[aps,prl,preprint,tightenlines,superscriptaddress,showpacs,byrevtex]{revtex4}
\usepackage{dcolumn}  

\usepackage{epsfig}
\newcommand{\mpp}{M_{p\bar{p}}}
\newcommand{\mb}{{M_{\rm bc}}}
\newcommand{\de}{{\Delta{E}}}
\newcommand{\bp}{{B^{+}}}
\newcommand{\bz}{{B^{0}}}
\newcommand{\pp}{{p\bar{p}}}
\newcommand{\ppk}{{p\bar{p}K^+}}
\newcommand{\pppi}{{p\bar{p}\pi^+}}
\newcommand{\ppkz}{{p\bar{p}K^0}}
\newcommand{\ppks}{{p\bar{p}K_S^0}}
\newcommand{\ppkst}{{p\bar{p}K^{* +}}}
\newcommand{\ppksz}{{p\bar{p}K^{*0}}}

\newcommand{\kst}{{K^{*{+}}}}
\newcommand{\ksz}{{K^{*0}}}
\newcommand{\ks}{{K_S^0}}

\begin{document}
\preprint{\vbox{ \hbox{   }
                  \hbox{Belle preprint 2003-23}
}}
\title{ \quad\\[0.5cm]
\boldmath Observation of $\bp \to \pppi$, $\bz \to \ppkz$, and
$\bp \to \ppkst$}


\affiliation{Budker Institute of Nuclear Physics, Novosibirsk}
\affiliation{Chiba University, Chiba}
\affiliation{University of Cincinnati, Cincinnati, Ohio 45221}
\affiliation{University of Frankfurt, Frankfurt}
\affiliation{Gyeongsang National University, Chinju}
\affiliation{University of Hawaii, Honolulu, Hawaii 96822}
\affiliation{High Energy Accelerator Research Organization (KEK), Tsukuba}
\affiliation{Hiroshima Institute of Technology, Hiroshima}
\affiliation{Institute of High Energy Physics, Chinese Academy of Sciences, Beijing}
\affiliation{Institute of High Energy Physics, Vienna}
\affiliation{Institute for Theoretical and Experimental Physics, Moscow}
\affiliation{J. Stefan Institute, Ljubljana}
\affiliation{Kanagawa University, Yokohama}
\affiliation{Korea University, Seoul}
\affiliation{Kyungpook National University, Taegu}
\affiliation{Institut de Physique des Hautes \'Energies, Universit\'e de Lausanne, Lausanne}
\affiliation{University of Ljubljana, Ljubljana}
\affiliation{University of Maribor, Maribor}
\affiliation{University of Melbourne, Victoria}
\affiliation{Nagoya University, Nagoya}
\affiliation{Nara Women's University, Nara}
\affiliation{National Lien-Ho Institute of Technology, Miao Li}
\affiliation{Department of Physics, National Taiwan University, Taipei}
\affiliation{H. Niewodniczanski Institute of Nuclear Physics, Krakow}
\affiliation{Nihon Dental College, Niigata}
\affiliation{Niigata University, Niigata}
\affiliation{Osaka City University, Osaka}
\affiliation{Osaka University, Osaka}
\affiliation{Panjab University, Chandigarh}
\affiliation{Peking University, Beijing}
\affiliation{Princeton University, Princeton, New Jersey 08545}
\affiliation{RIKEN BNL Research Center, Upton, New York 11973}
\affiliation{Saga University, Saga}
\affiliation{University of Science and Technology of China, Hefei}
\affiliation{Seoul National University, Seoul}
\affiliation{Sungkyunkwan University, Suwon}
\affiliation{University of Sydney, Sydney NSW}
\affiliation{Tata Institute of Fundamental Research, Bombay}
\affiliation{Toho University, Funabashi}
\affiliation{Tohoku Gakuin University, Tagajo}
\affiliation{Tohoku University, Sendai}
\affiliation{Department of Physics, University of Tokyo, Tokyo}
\affiliation{Tokyo Institute of Technology, Tokyo}
\affiliation{Tokyo Metropolitan University, Tokyo}
\affiliation{Tokyo University of Agriculture and Technology, Tokyo}
\affiliation{Toyama National College of Maritime Technology, Toyama}
\affiliation{University of Tsukuba, Tsukuba}
\affiliation{Utkal University, Bhubaneswer}
\affiliation{Virginia Polytechnic Institute and State University, Blacksburg, Virginia 24061}
\affiliation{Yokkaichi University, Yokkaichi}
\affiliation{Yonsei University, Seoul}
  \author{M.-Z.~Wang}\affiliation{Department of Physics, National Taiwan University, Taipei} 
  \author{K.~Abe}\affiliation{High Energy Accelerator Research Organization (KEK), Tsukuba} 
  \author{K.~Abe}\affiliation{Tohoku Gakuin University, Tagajo} 
  \author{T.~Abe}\affiliation{High Energy Accelerator Research Organization (KEK), Tsukuba} 
  \author{I.~Adachi}\affiliation{High Energy Accelerator Research Organization (KEK), Tsukuba} 
  \author{H.~Aihara}\affiliation{Department of Physics, University of Tokyo, Tokyo} 
  \author{Y.~Asano}\affiliation{University of Tsukuba, Tsukuba} 
  \author{T.~Aso}\affiliation{Toyama National College of Maritime Technology, Toyama} 
  \author{V.~Aulchenko}\affiliation{Budker Institute of Nuclear Physics, Novosibirsk} 
  \author{T.~Aushev}\affiliation{Institute for Theoretical and Experimental Physics, Moscow} 
% ...

\collaboration{The Belle Collaboration}


\begin{abstract}
We report the first observation of a $b\to u$ type charmless
baryonic $B$ decay, $\bp \to \pppi$, as well as $b\to s$ type $\bz
\to \ppkz$ and $\bp \to \ppkst$ decays. The analysis is based on a
78~fb$^{-1}$ data sample recorded on the $\Upsilon({\rm 4S})$
resonance with the Belle detector at KEKB.
We find ${\mathcal B}(\bp \to \pppi) = ( 3.06^{+0.73}_{-0.62} \pm
0.37) \times 10^{-6}$, ${\mathcal B}(B^0 \to \ppkz) =
(1.88^{+0.77}_{-0.60} \pm 0.23) \times 10^{-6}$, and ${\mathcal
B}(\bp \to \ppkst) = ( 10.3^{+3.6 + 1.3}_{-2.8 -1.7} ) \times
10^{-6}$. 
We also update ${\mathcal B}(\bp \to \ppk) = (5.66^{+0.67}_{-0.57}
\pm 0.62 )\times 10^{-6}$, and present an upper limit on
${\mathcal B}(\bz \to \ppksz)$ at the 90\% confidence level.
%
A common feature of the observed decay modes is threshold peaking in
baryon pair invariant mass.
\end{abstract}
\maketitle
\tighten
{\renewcommand{\thefootnote}{\fnsymbol{footnote}}
\setcounter{footnote}{0}

The Belle Collaboration recently reported the observation of $\bp
\to \ppk$~\cite{ppk} and $\bz \to p\bar \Lambda\pi^-$~\cite{plpi}
decays. These are the first examples of $B$ meson decays to
charmless three-body final states containing baryons, and are
candidates for  $b\to s$ penguin transitions. 
Our observation of these modes has stimulated much theoretical 
interest~\cite{CY,CHT,suzuki,newpa,glueball,rus,RosnerB} 
due to these decay 
channels may be used to 
test our theoretical understanding of
rare decay processes involving
baryons and search for
direct $CP$ violation.
One interesting feature of the observation is that 
the baryon pair mass 
spectra seem to peak toward threshold as originally conjectured 
by Refs.~\cite{HS,rhopn}. Lately more speculations on this are 
proposed~\cite{glueball,rus,RosnerB}. 
In this letter we report the first observation 
of $\bp \to \pppi$~\cite{conjugate}, which is dominated by the $b\to u$ tree
diagram. We also report the first observation of $\bz \to \ppkz$
and $\bp \to \ppkst$ decays, and improve the measurement of $\bp
\to \ppk$. A search for the $\bz \to \ppksz$ mode yields only an
upper limit.
With these new results we study the $\pp$ mass spectra to see 
if the threshold peaking observed in our previous results is confirmed 
with the newly observed modes.

%

We use a  78 fb$^{-1}$  data sample,
consisting of $85.0 \pm 0.5$ million $B\bar{B}$ pairs,
collected by the Belle detector 
at the KEKB asymmetric energy $e^+e^-$ (3.5 on 8~GeV) collider~\cite{KEKB}.
The Belle detector is a large solid angle magnetic spectrometer
that consists of a three layer silicon vertex detector (SVD), a 50
layer central drift chamber (CDC), an array of aerogel threshold
\v{C}erenkov counters (ACC), a barrel-like arrangement of time of
flight scintillation counters (TOF), and an electromagnetic
calorimeter comprised of CsI (Tl) crystals located inside a
superconducting solenoid coil that provides a 1.5~T magnetic
field.  An iron flux return located outside of the coil is
instrumented to detect $K_L^0$ mesons and to identify muons. The
detector is described in detail elsewhere~\cite{Belle}.

The event selection criteria are based on the information obtained
from the tracking system
(SVD+CDC) and the hadron identification system (CDC+ACC+TOF),
and are optimized using Monte Carlo (MC)
simulated event samples.
All primary charged tracks
are required to satisfy track quality criteria
based on the track impact parameters relative to the
interaction point (IP). 
The deviations from the IP position are required to be within
$\pm$1 cm in the transverse ($x$--$y$) plane, and within $\pm$3 cm
in the $z$ direction, where the $z$ axis is defined by the
positron beam line. Proton, kaon and pion candidates are selected 
using $p/K/\pi$ likelihood functions obtained from the hadron
identification system. For protons, we require $L_p/(L_p+L_K)> 0.6 $ and
$L_p/(L_p+L_{\pi})> 0.6$, where $L_{p/K/\pi}$ stands for the
proton/kaon/pion likelihood.
We require $L_K/(L_K+L_{\pi})> 0.6$ to identify kaons and  
$L_{\pi}/(L_K+L_{\pi})> 0.6$ for pions.
$\ks$
candidates are reconstructed via the $\pi^+\pi^-$ decay channel
and have an invariant mass with  $ |M_{\pi^+\pi^-} - M_{K^0}| < 30$ MeV/$c^2$.
The candidate must have a displaced vertex and flight direction consistent with
a $\ks$ originating from the interaction point. 
We use the selected kaons and pions to form $\kst$
($\to \ks\pi^+$) and  $\ksz$ ($\to K^+\pi^-$) candidates by requiring the
invariant mass of the combination to be within 80 MeV/$c^2$ of the
$K^*$ mass. The numbers used are 892 MeV/$c^2$ and
896 MeV/$c^2$ for $M_{\kst}$ and $M_{\ksz}$, respectively.
To ensure the decay process is
genuinely charmless, we apply the charm veto. 
The regions  $2.85<M_{p \bar{p}}<3.128$ GeV/$c^2$
and $3.315<M_{p \bar{p}}<3.735$ GeV/$c^2$ are excluded to remove background 
from modes with $\eta_c, J/\psi$ and 
$\psi^{\prime},\chi_{c0},\chi_{c1}$ mesons, respectively.
The region  $2.262<M_{p \ks}<2.310$ GeV/$c^2$
is excluded to remove $\Lambda_c$ candidates.

Since the initial collision 
center of mass energy is set to match the $\Upsilon({\rm
4S})$ resonance, which decays into a $B\bar{B}$ pair, one can use
the following two kinematic variables to identify the
reconstructed $B$ meson candidates: the beam
constrained mass $\mb = \sqrt{E^2_{\rm beam}-p^2_B}$, and the
energy difference $\de = E_B - E_{\rm beam}$, where $E_{\rm
beam}$, $p_B$, and $E_B$ are the beam energy, the momentum, and
energy of the reconstructed $B$ meson, respectively, in the rest frame of
the $\Upsilon({\rm 4S})$. The candidate region is
defined as 5.2 GeV/$c^2 < \mb < 5.29$ GeV/$c^2$ and $|\de|< 0.2$
GeV.

The dominant background arises from the continuum $e^+e^-
\to q\bar{q}$ process, 
with much smaller contributions from ``cross-feed'', where similar 
types of rare decay events pass each other's signal criteria.
The background from charm-bearing and charmless mesonic decays is 
negligible.
In the $\Upsilon({\rm 4S})$ rest frame,
continuum events are jet-like while
$B\bar{B}$ events are more spherical. 
One can use the reconstructed momenta 
of final state particles to form various shape variables (e.g. thrust
angle, Fox-Wolfram moments, etc.) in order to categorize each event.  
We follow the scheme defined in Ref.~\cite{etapk} that
combines seven event shape variables into
a Fisher discriminant~\cite{fisher} in order to suppress
continuum background. The variables chosen have
almost no correlation
with $\mb$ and $\de$.
Probability density functions (PDFs) for the Fisher discriminant and
the cosine of the angle between the $B$ flight direction
and the beam direction in the $\Upsilon({\rm 4S})$ rest frame
are combined to form the signal (background)
likelihood ${\cal L}_{\rm S (BG)}$.
The signal PDFs are determined using signal MC
simulation; the background PDFs are obtained from the continuum MC
simulation for events with
5.27 GeV/$c^2 < \mb < $5.29 GeV/$c^2$
 and $|\de|< 0.1$ GeV.
We require
the likelihood ratio ${\cal LR} = {\cal L}_{\rm S}/({\cal L}_{\rm
S}+{\cal L}_{\rm BG})$ to be greater than 0.8, 0.9, 0.85, 0.8 and 0.8 for
$\ppk$, $\pppi$, $\ppks$, $\ppkst$ and $\ppksz$ modes, respectively.
The selection
points are determined by optimization of $S/\sqrt{(S+B)}$, where $S$ and $B$
denote the number of signal and background, respectively. Note that
a nominal signal branching fraction of $4 \times 10^{-6}$ is assumed.

\begin{figure}[p]
\centering 
\epsfig{file=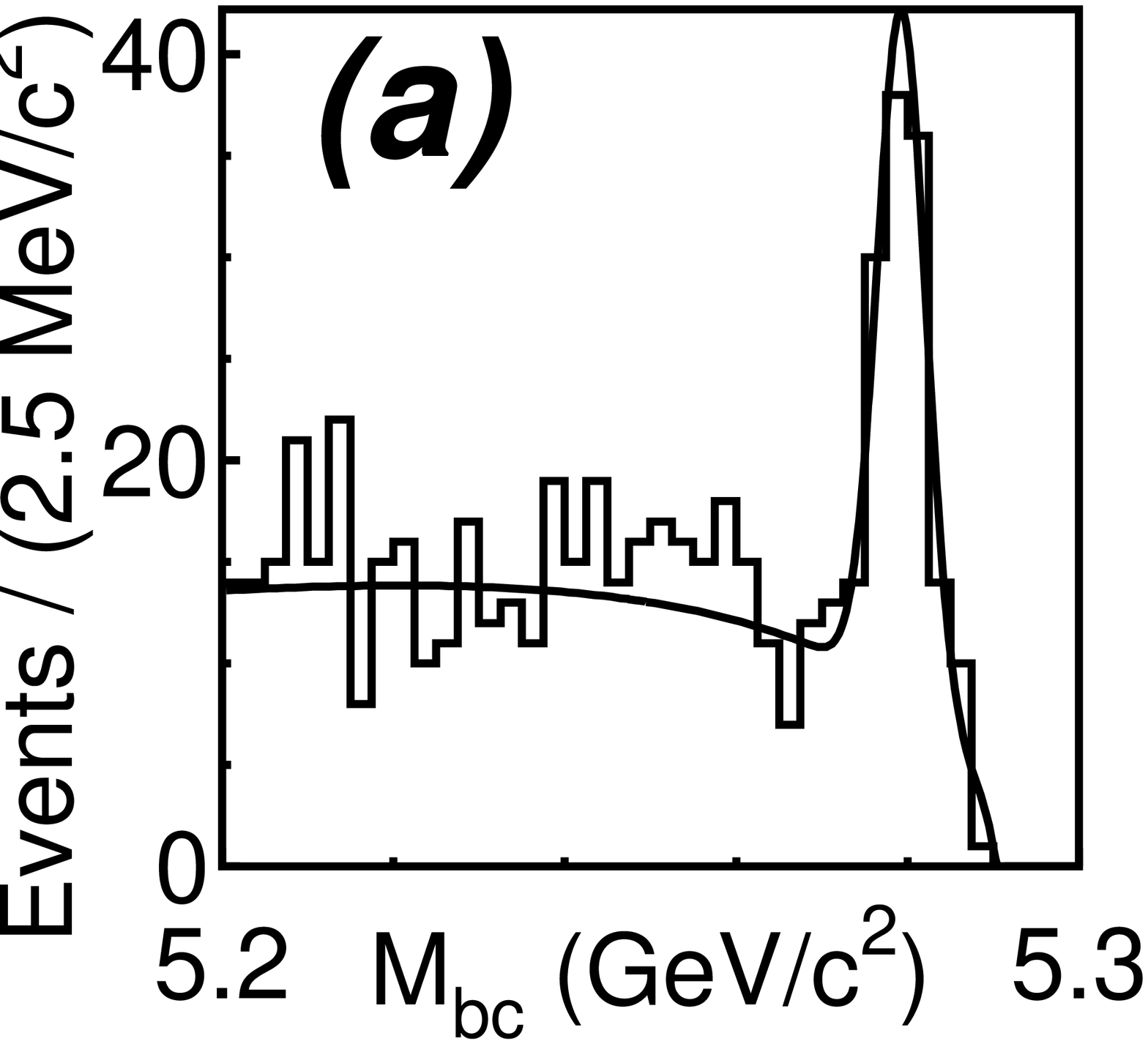,width=5cm,height=4.3cm}
\hskip 1cm 
\epsfig{file=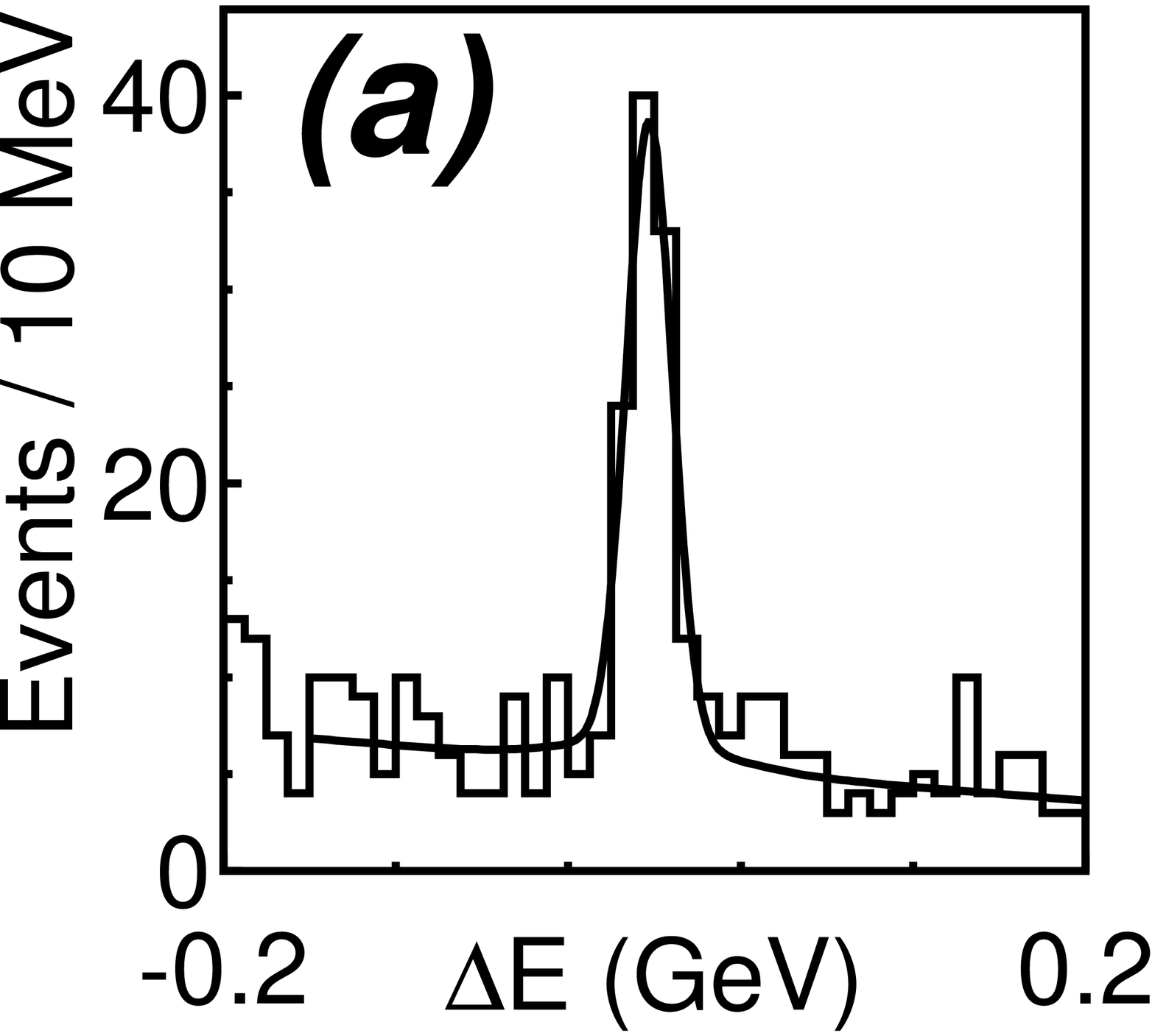,width=5cm,height=4.3cm} 
\hskip 1cm
\epsfig{file=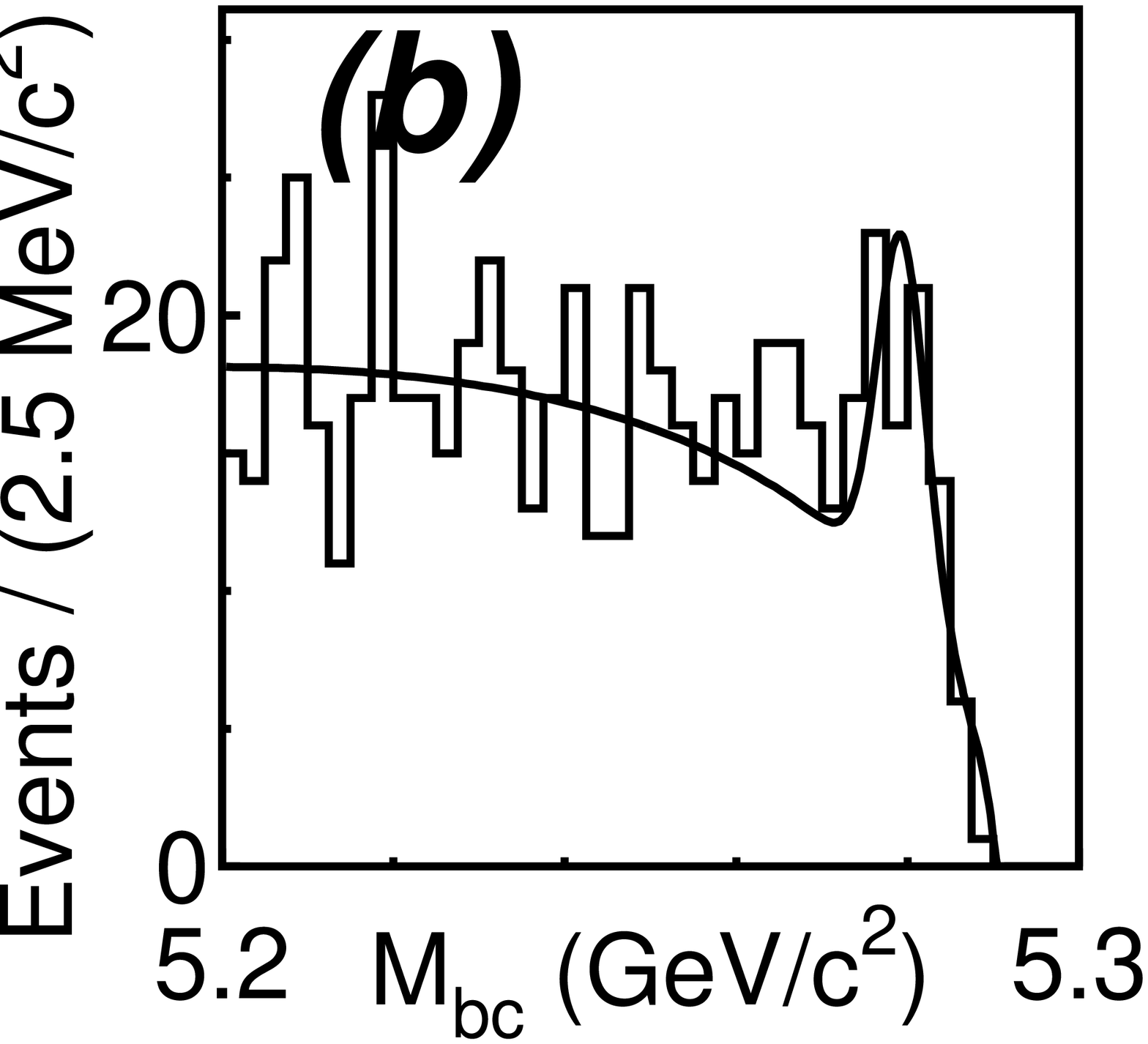,width=5cm,height=4.3cm} 
\hskip 1cm 
\epsfig{file=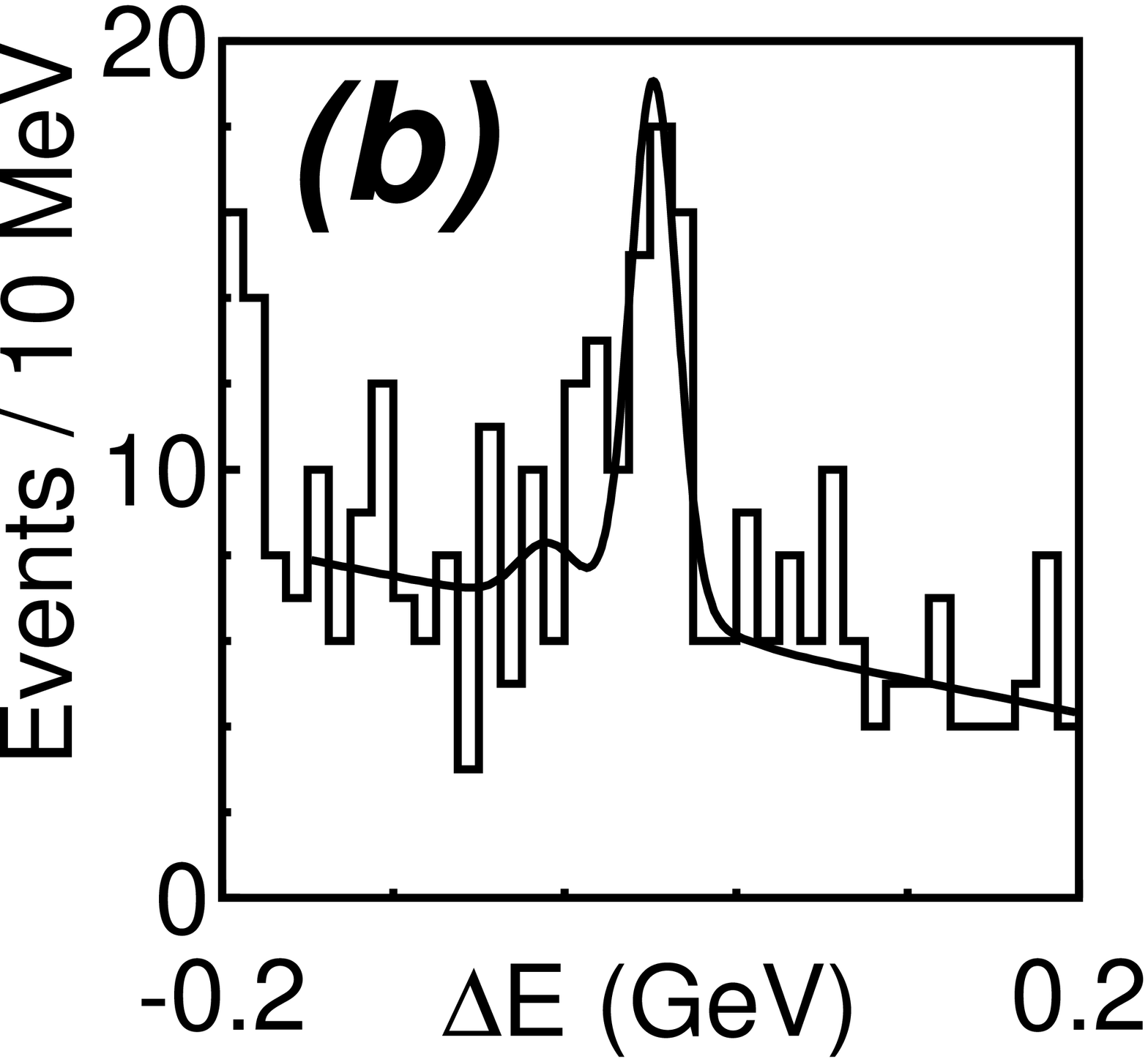,width=5cm,height=4.3cm} 
\hskip 1cm 
\epsfig{file=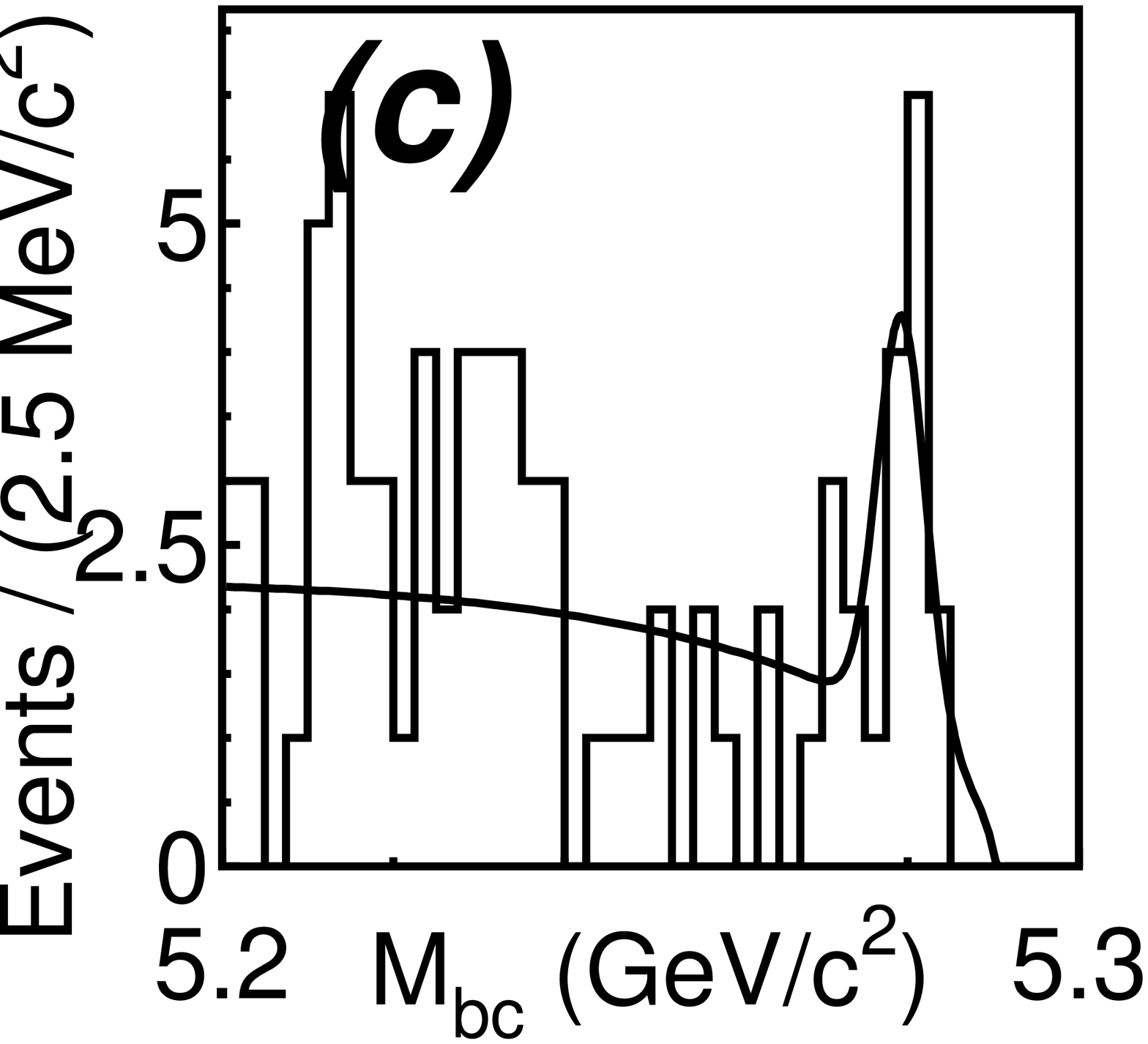,width=5cm,height=4.3cm} 
\hskip 1cm 
\epsfig{file=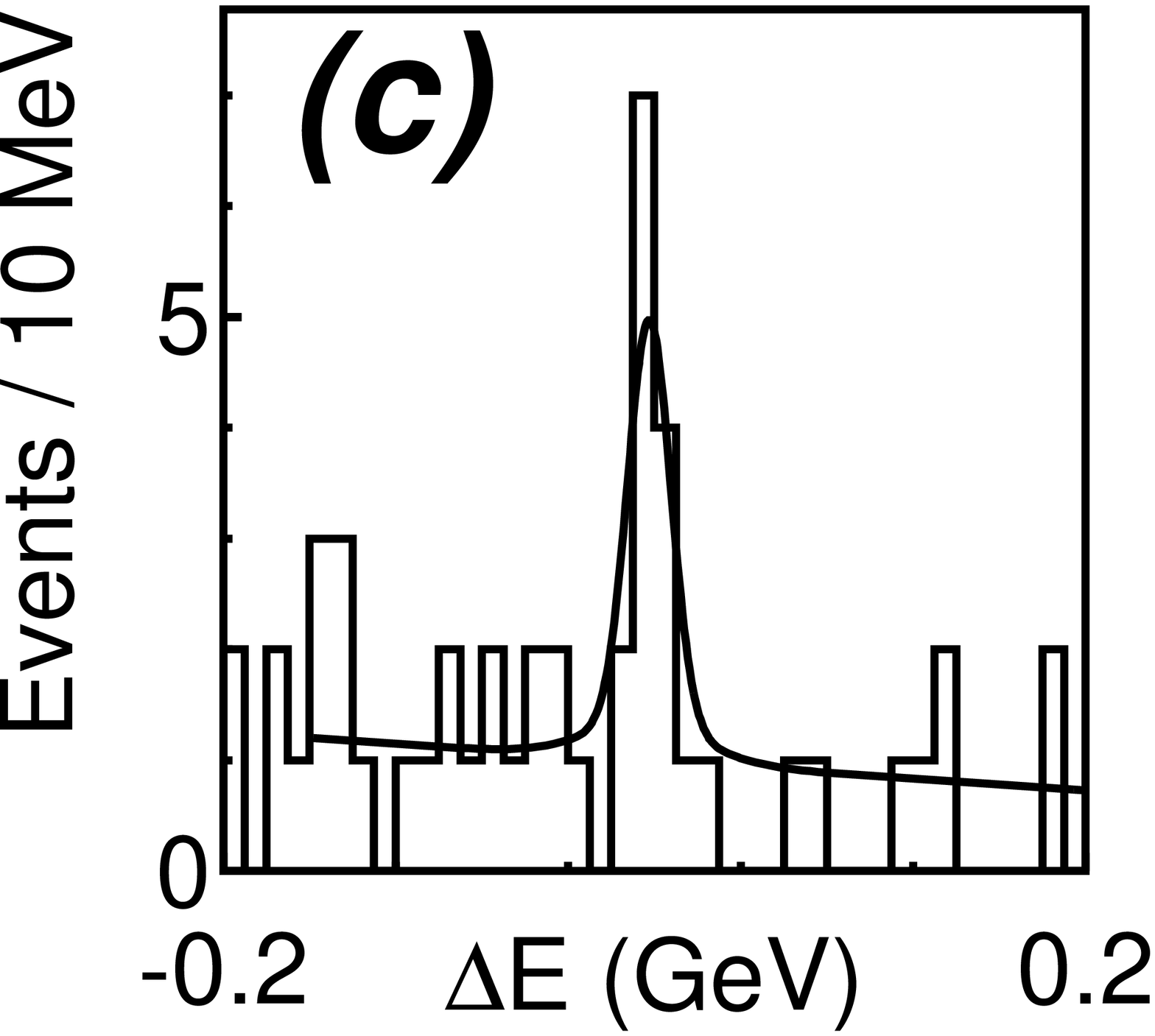,width=5cm,height=4.3cm} 
\hskip 1cm 
\epsfig{file=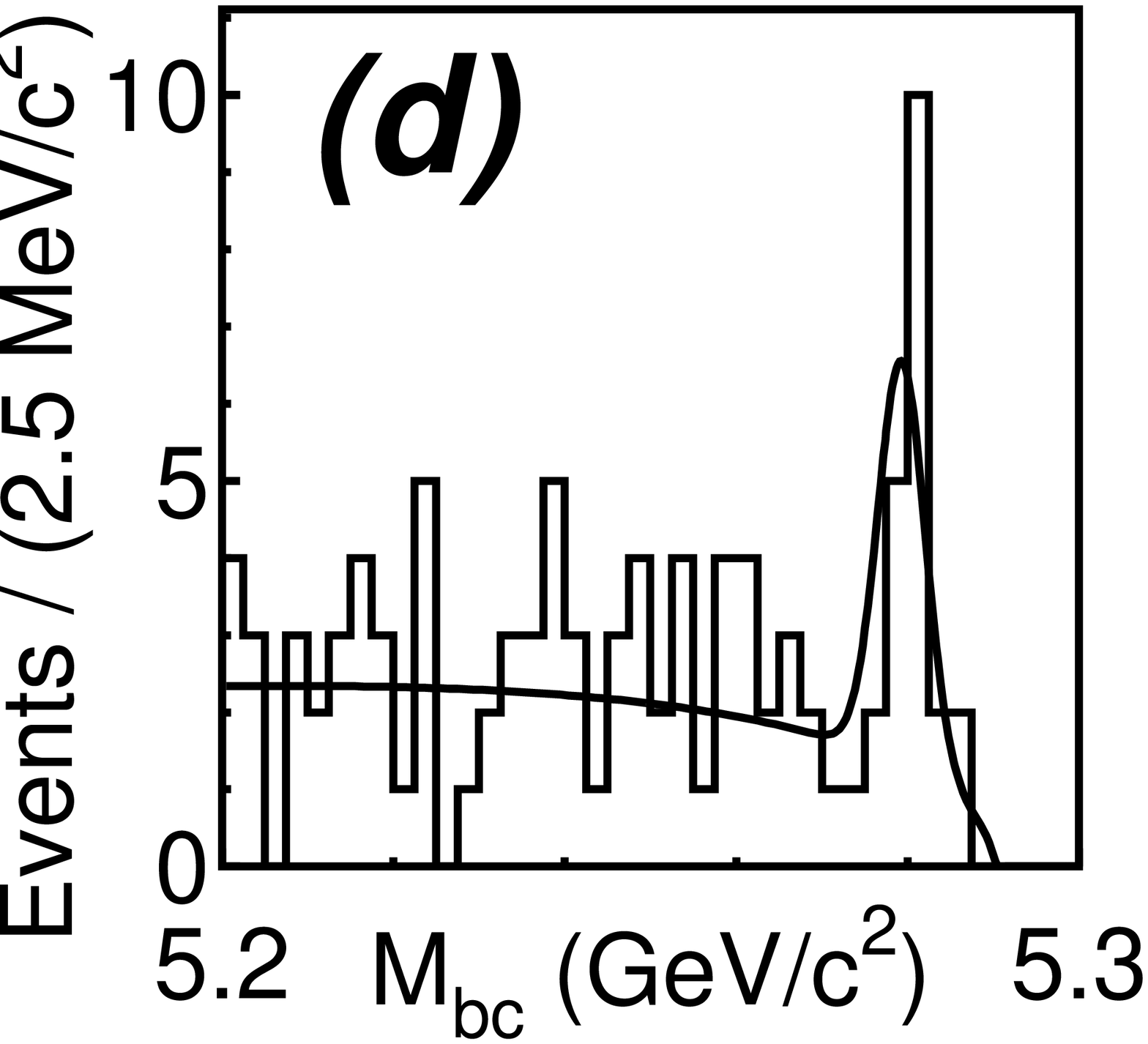,width=5cm,height=4.3cm} 
\hskip 1cm 
\epsfig{file=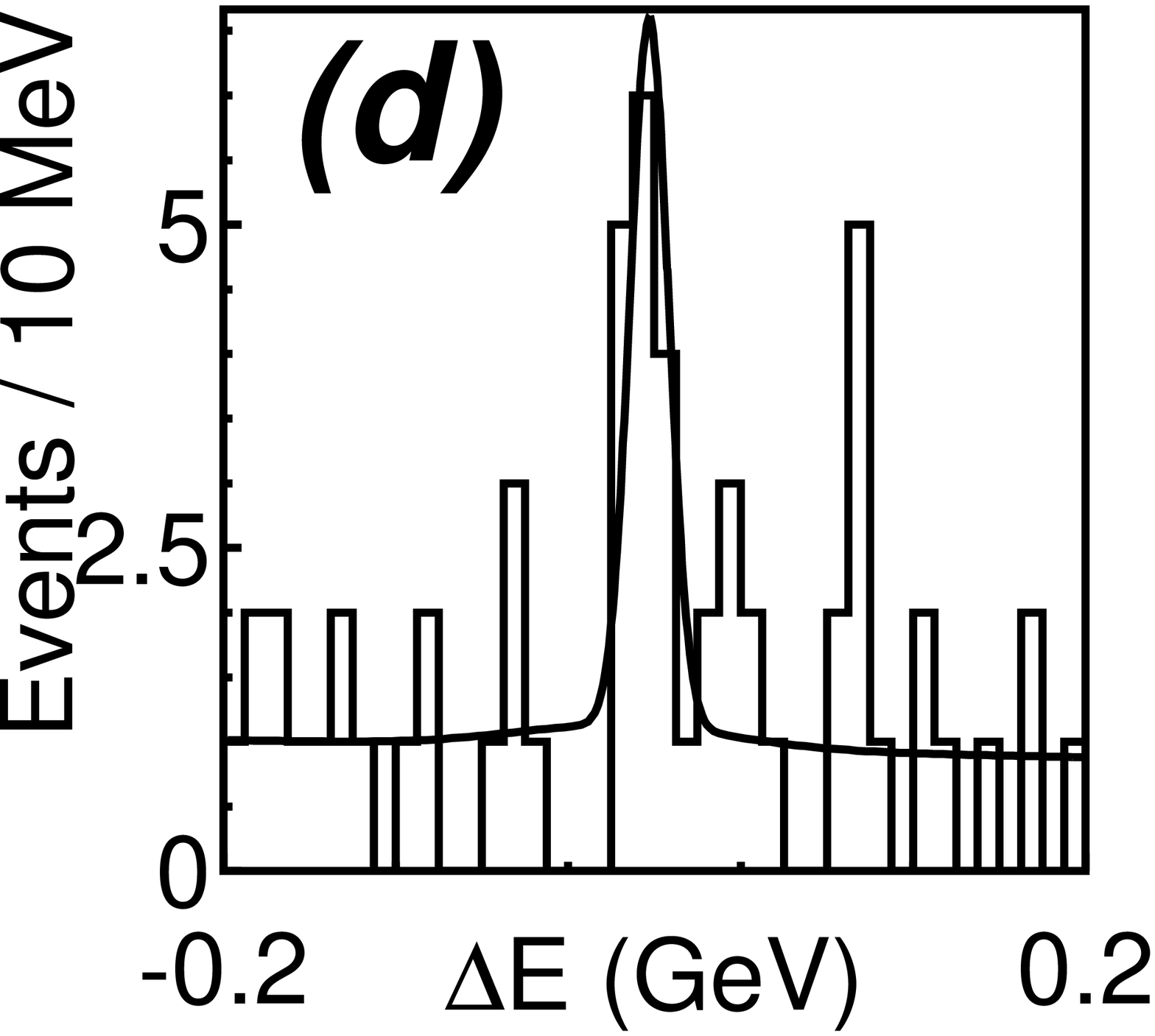,width=5cm,height=4.3cm} 
\hskip 1cm 
\centering
\caption{ $\mb$ and $\de$ distributions for (a) $\ppk$, (b)
$\pppi$, (c) $\ppks$, and (d) $\ppkst$ 
modes, for
$M_{p\bar p} < 2.85$ GeV/$c^2$. 
}

\label{fg:mergembde}
\end{figure}

\begin{figure}[b!]
\centering
\mbox{\psfig{figure=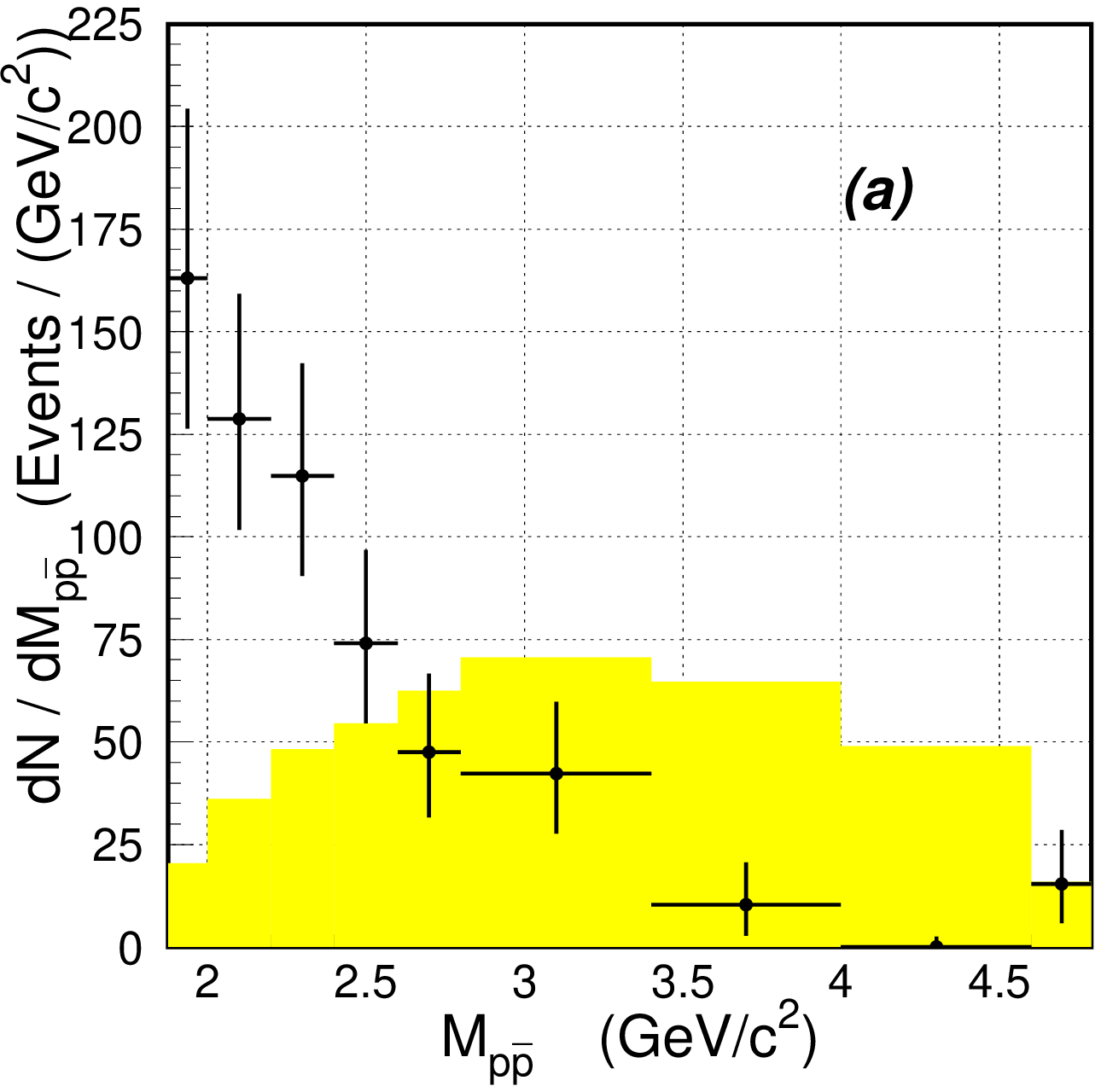,width=2.5in}}
\mbox{\psfig{figure=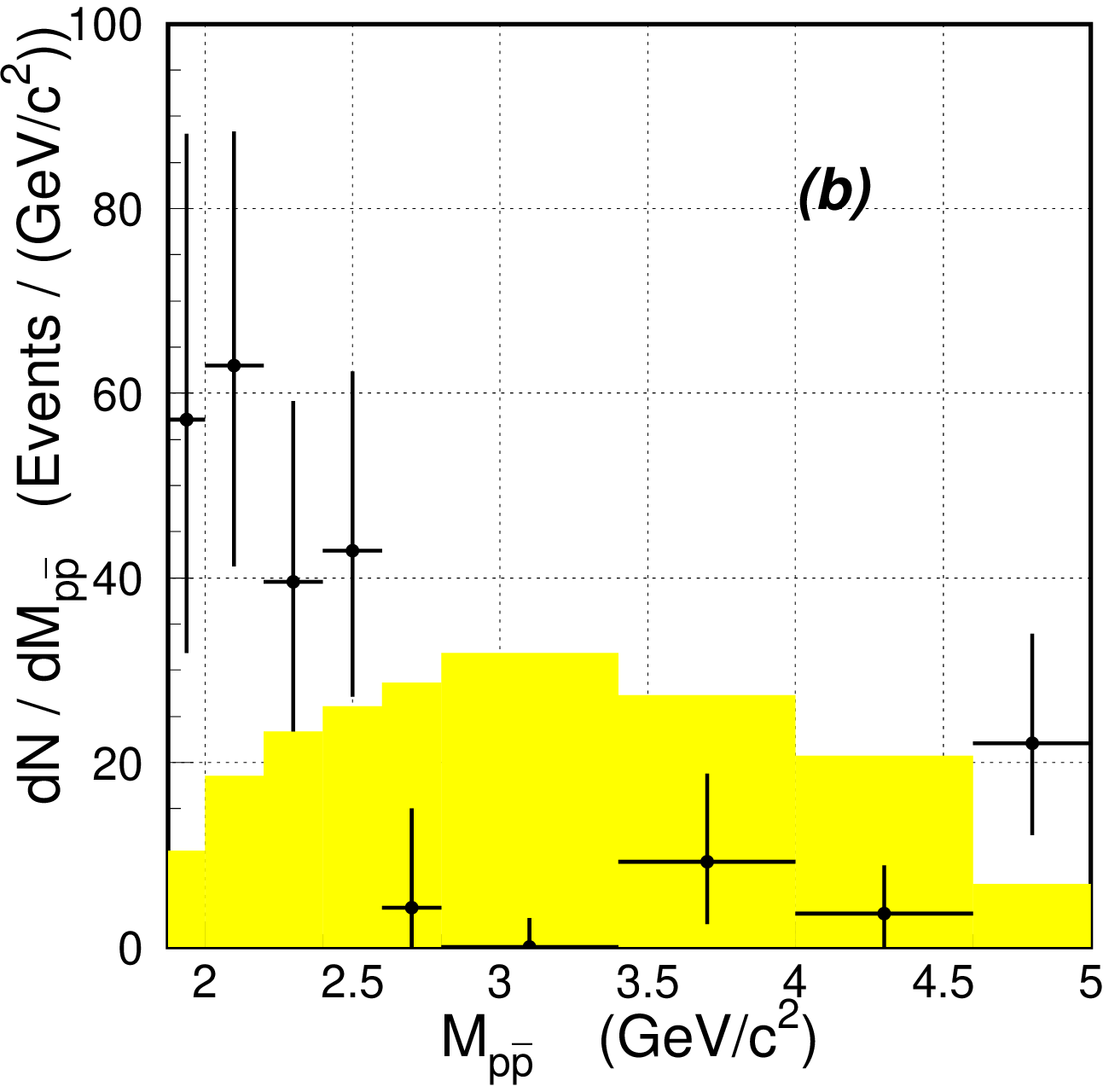,width=2.5in}}
\mbox{\psfig{figure=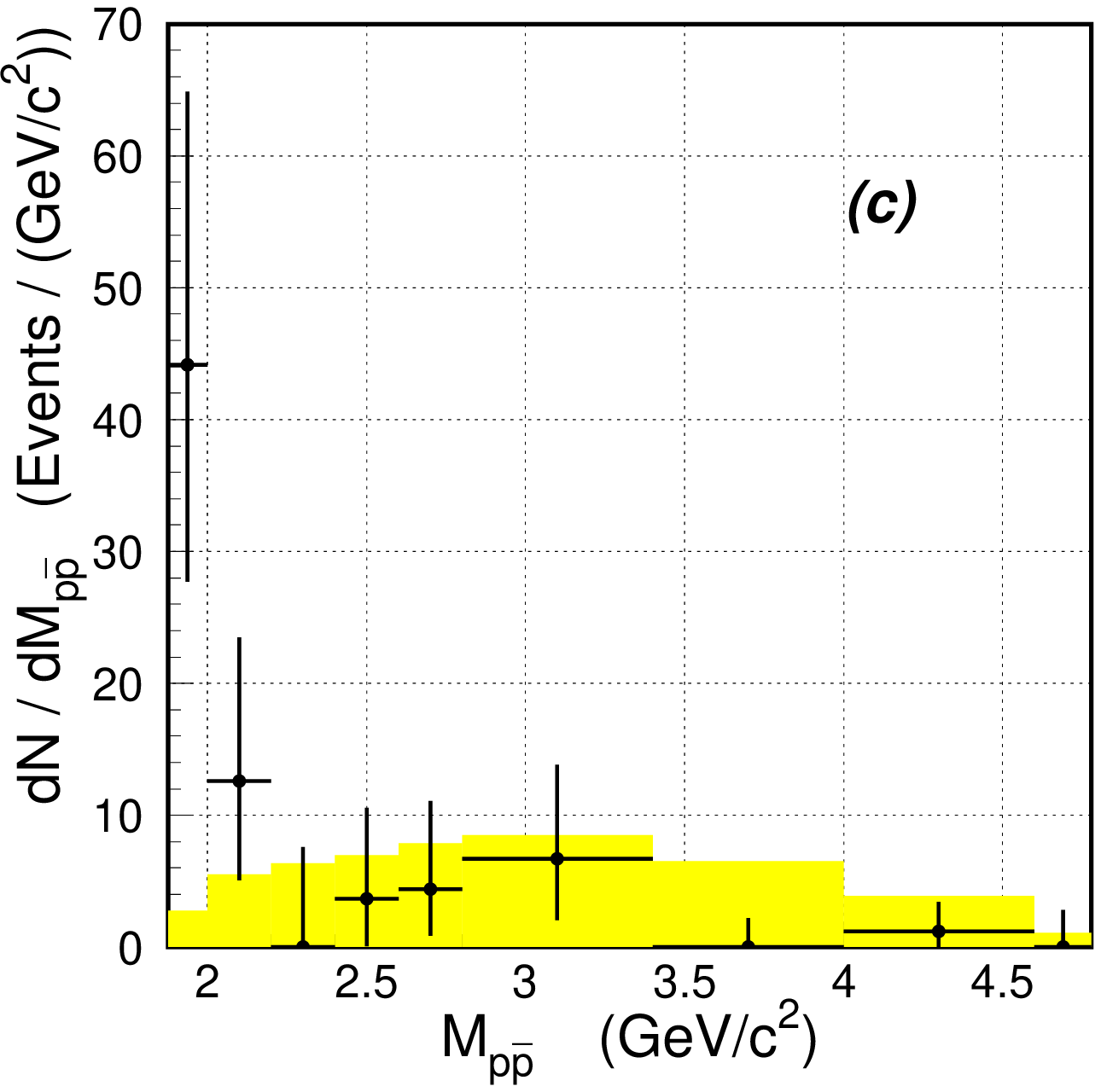,width=2.5in}}
\mbox{\psfig{figure=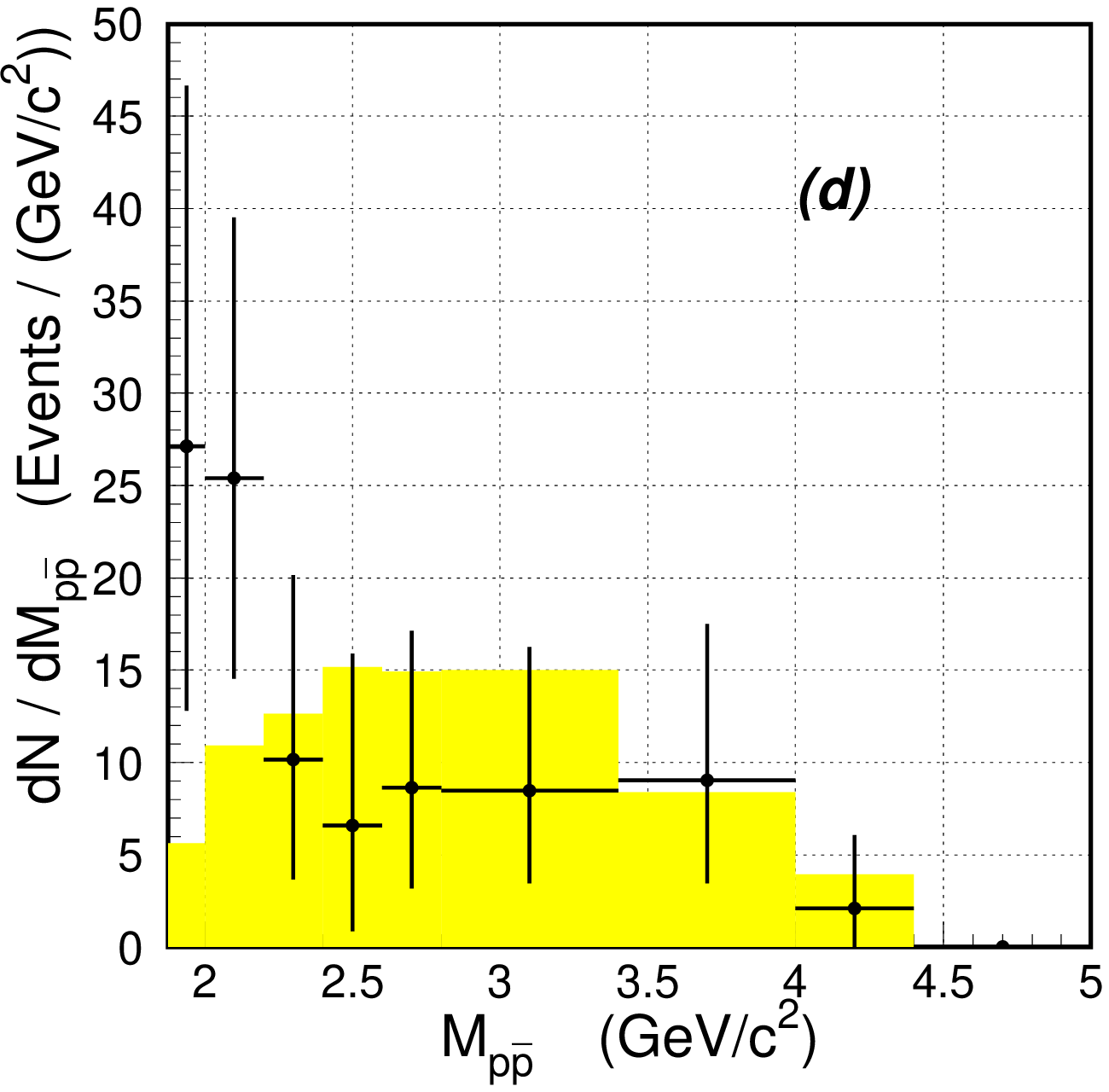,width=2.5in}}
\centering
\caption{Fitted signal yield divided by the bin size for
(a) $\ppk$, (b) $\pppi$, (c) $\ppks$, and (d) $\ppkst$ 
modes in bins of $\mpp$. The shaded distribution is from 
the phase-space MC simulation with area normalized to
signal yield.
}
\label{fg:allphase}
\end{figure}

Since the previous studies~\cite{ppk,plpi} showed an enhancement at
low baryon-antibaryon invariant mass, we first focus on the near
threshold region by 
requiring $\mpp < 2.85$ GeV/$c^2$ to make sure it
is below charmonium threshold.
The $\mb$ distributions (with $|\de|<$ 0.05
GeV), and the $\de$ distributions (with $\mb >$ 5.27 GeV/$c^2$)
for the $\ppk$, $\pppi$, $\ppks$, and $\ppkst$ 
modes are
shown in Fig.~\ref{fg:mergembde}. We use an unbinned likelihood fit
to estimate the signal yield:
$$ L = \prod_{i=1}^{N} [f_sP_s(M_{{\rm bc}_i},\Delta{E}_i)+
(1-f_s)P_b(M_{{\rm bc}_i},\Delta{E}_i)],$$
where $P_s(P_b)$ denotes the signal (background) PDF and $f_s$ is the signal
fraction of the total $N$ candidates.
For the signal PDF,
we use a Gaussian in $\mb$ and a double Gaussian in $\de$.  We fix
the parameters of these functions to values determined by MC simulation.
Background shapes are studied using sideband
events: 0.1 GeV $ < |\de| < 0.2$ GeV for the $\mb$ study and
5.20 GeV/$c^2$ $ < \mb <$ 5.26 GeV/$c^2$ for $\de$.
These shapes are confirmed with a continuum MC sample.
We use the following parametrization first used by the ARGUS collaboration, 
$ f(\mb)\propto \mb\sqrt{1-x^2}
\exp[-\xi (1-x^2)]$,  
to model
the $\mb$ background, where $x$ is defined as $\mb/E_{\rm beam}$ and $\xi$ is
a parameter to be fit. 
The $\de$ background shape is modeled by a first order polynomial.
There are possible cross-feeds from
$\ppkst$ and $\ppksz$ modes to $\ppk$, $\pppi$ and $\ppks$
modes; therefore the cross-feed region ($\de < -0.16$ GeV) 
is excluded in the fit.
Since the $\pppi$ mode can contain
non-negligible cross-feed events from the $\ppk$ mode, we include the
$\ppk$ MC cross-feed shape in the fit for the determination of the
$\pppi$ yield. 
The fit results are shown in 
Fig.~\ref{fg:mergembde} by solid curves. 
The fit yields are
96.4 $^{+11.2}_{-10.5}$,
37.4 $^{+8.1}_{-7.7}$,
11.3 $^{+4.1}_{-3.4}$,
and 14.5 $^{+4.6}_{-4.0}$
with
significances of $15.3$, $6.7$, $5.1$, and $6.0$ standard
deviations for
the $\ppk$, $\pppi$, $\ppks$, and $\ppkst$ modes, respectively. The 
significance is defined as $\sqrt{-2 {\rm ln}(L_0/L_{max})}$~\cite{PDG}, 
where $L_0$ and
$L_{max}$ denote the likelihood with signal yield fixed at
zero and at the fitted value, respectively. 

\begin{table}[b]
\caption{Branching fractions (in units of $10^{-6}$) 
in different $\mpp$ bins for the
$\ppk$, $\pppi$, $\ppks$ and $\ppkst$ modes.} \label{br}
\begin{center}
\begin{tabular}{ccccc}
$\mpp$ (GeV/$c^2$)&
$\ppk$&$\pppi$&$\ppks$&$\ppkst$
\\
\hline $1.876-2.0$&$0.91^{+0.23}_{-0.20}$ \ &
$0.29^{+0.15}_{-0.13}$ \ & $0.38^{+0.18}_{-0.14}$ \ & $1.40
^{+1.01}_{-0.74}$
\\
\hline $2.0-2.2$ \ & $1.33^{+0.31}_{-0.28}$ \ &
$0.57^{+0.23}_{-0.20}$ \ & $0.17^{+0.15}_{-0.10}$ \ & $2.11
^{+1.17}_{-0.90}$
\\
\hline $2.2-2.4$ \ &$1.20^{+0.29}_{-0.25}$ \ &
$0.39^{+0.19}_{-0.16}$ \ & $0.00^{+0.12}_{-0.12}$ \ & $1.01
^{+1.00}_{-0.65}$
\\
\hline $2.4-2.6$ \ & $0.82^{+0.25}_{-0.22}$ \ &
$0.46^{+0.21}_{-0.17}$ \ & $0.06^{+0.12}_{-0.06}$ \ & $0.64
^{+0.91}_{-0.56}$
\\
\hline $2.6-2.8$ \ & $0.51^{+0.21}_{-0.17}$ \ &
$0.05^{+0.12}_{-0.12}$ \ & $0.08^{+0.12}_{-0.06}$ \ & $0.95
^{+0.94}_{-0.60}$
\\
\hline $2.8-3.4$ \ & $0.53^{+0.22}_{-0.18}$ \ &
$0.00^{+0.04}_{-0.04}$ \ & $0.14^{+0.15}_{-0.10}$ \ & $1.19
^{+1.09}_{-0.71}$
\\
\hline $3.4-4.0$ \ & $0.15^{+0.15}_{-0.11}$ \ &
$0.15^{+0.16}_{-0.11}$ \ & $0.00^{+0.07}_{-0.07}$ \ & $2.07
^{+1.93}_{-1.28}$
\\
\hline $4.0-4.6$ \ & $0.00^{+0.08}_{-0.08}$ \ &
$0.15^{+0.21}_{-0.21}$ \ & $0.10^{+0.19}_{-0.10}$ \ & $0.93
^{+1.75}_{-1.75}$
\\
\hline $4.6-M_{p\bar{p}-{\rm lim}}$ \ & $0.20^{+0.17}_{-0.12}$ \ &
$1.00^{+0.54}_{-0.45}$ \ & $0.00^{+0.11}_{-0.11}$ \ & n/a
\\
\end{tabular}
\end{center}
\end{table}

Although we do not have adequate statistics to perform a full Dalitz
plot analysis, the observed distribution is not uniform over phase
space. To reduce the model dependence
in determining the branching
fraction, we fit the signal yields separately 
in nine bins of $\mpp$, and correct for the detection efficiencies
from MC simulation in each bin. In Fig.~\ref{fg:allphase}, we show
the signal yield versus $M_{p\bar p}$, with three-body phase space
from MC (normalized in area) superimposed. The observed mass
distributions all peak at low $p\bar{p}$ mass. The branching
fractions (${\cal B}$) in bins of $M_{p\bar{p}}$ for the
observed modes are given in Table~\ref{br}. 
The upper limit of the last bin is different for each mode and is 
equal to the kinematic limit, $M_{p\bar p-{\rm lim}}$.
We sum the partial
branching fractions to obtain the total branching fractions.
The results are listed in Table~\ref{tab:tot} and
the  branching fractions below charmonium threshold, $M_{p\bar p} < 2.85$
GeV/$c^2$, are also listed for comparison. Note that 
${\cal B}(\bz \to \ppkz) = 2{\cal B}(\bz \to \ppks)$ is
assumed.

The search for $\ppksz$ gives a yield of 13$^{+6}_{-5}$ events with
  a significance of about 3 standard deviations. 
Since it is less significant, 
we use the fit results to estimate the expected background
and compare this  with the observed number of events
in the signal region
in order to set the upper limit on the
yield at the 90\% confidence level~\cite{Gary,Conrad}.
Note that the systematic uncertainty is needed in this estimation.
The upper limit yield of the $\ppksz$ mode in the full
$\mpp$ range is determined to
be 57 at the 90\% confidence level. The
branching fraction is found to be
${\cal B}(\bz \to \ppksz) < 7.6 \times 10^{-6}$.

\begin{table}[htb]
\caption{Branching fractions (in units of $10^{-6}$) 
in the full $\mpp$ range and below the 
charm threshold ($M_{p\bar p} < 2.85$GeV/$c^2$)
for the
$\ppk$, $\pppi$, $\ppks$ and $\ppkst$ modes. Statistical and systematic errors
are quoted.}
\label{tab:tot} 
\begin{center}
\begin{tabular}{ccc}
mode& full $\mpp$ range & $M_{p\bar p} < 2.85$GeV/$c^2$ \\
\hline
$\bp \to \ppk$& 
$5.66^{+0.67}_{-0.57} \pm 0.62 $&
$4.89^{+0.59}_{-0.55} \pm 0.54 $ \\
\hline
$\bp \to \pppi$&
$3.06^{+0.73}_{-0.62} \pm 0.37 $&
$1.76^{+0.42}_{-0.37} \pm 0.21 $ \\
\hline
$\bz \to \ppkz$&
$1.88^{+0.77}_{-0.60}  \pm 0.23 $&
$1.56^{+0.52}_{-0.49}  \pm 0.19 $ \\
\hline
$\bp \to \ppkst$&
$10.3^{+3.6 + 1.3}_{-2.8 - 1.7}$&
$6.7^{+2.4 + 0.9}_{-2.0 - 1.1} $
\\
\end{tabular}
\end{center}
\end{table}

\vskip 0.3cm

Systematic uncertainties 
are studied using high statistics control samples. For proton
identification, we use a  $\Lambda \to p \pi^-$ sample, while for
$K/\pi$ identification we use a $D^{*+} \to D^0\pi^+$,
 $D^0 \to K^-\pi^+$ sample.
Tracking efficiency is studied with
fully and partially reconstructed $D^*$ samples.
$\ks$ reconstruction efficiency is studied with a $D^- \to \ks\pi^-$
sample.
The $\cal LR$ continuum suppression uncertainty is studied with 
$\bp \to J/\psi K^+$, $\bz \to J/\psi\ks$, $\bp \to J/\psi\kst$, and
$\bz \to J/\psi\ksz$
(with $J/\psi \to \mu^+\mu^-$) control samples.
Based on these studies,
we assign a 1\% error for each track, 3\% for each proton identification,
2\% for each kaon/pion identification, 5\% for $\ks$ reconstruction
and 6\% for the $\cal LR$ selection.

The systematic uncertainty  in the fit yield is studied by varying
the parameters of the signal and background PDFs. We assign an
error of 5\% for the $\pppi$ mode and 4\% for the other modes. 
There is a  possibility of non-resonant $\ks \pi^+$
combinations passing our $\kst$ cut.  We study this by fitting for the $B$
yield in bins of $M_{\ks\pi^+}$.
Choosing a Breit--Wigner signal and a first order polynomial
background to estimate the $\bp \to \ppkst$ branching fraction,
the number drops by 9\%. 
We include this uncertainty in the estimation
of signal yield for the $\ppkst$ mode. The MC statistical
uncertainty and modeling with nine $\mpp$ bins contributes a 2\%
error in the branching fraction determination. The error on the
number
of $B\bar{B}$ pairs is determined to be 1\%, where an
assumption is made that the branching fractions of $\Upsilon({\rm 4S})$ 
to neutral and charged $B\bar{B}$ pairs are equal. 


We first sum the correlated errors linearly (e.g., total 5\%
tracking error for the $\ppkst$ mode) and then combine with the
uncorrelated ones in quadrature. The determined systematic
uncertainties are 11\%, 12\%, 12\%, $^{+13}_{-16}$\%, and 13\% for
the $\ppk$, $\pppi$, $\ppkz$, $\ppkst$, and $\ppksz$ modes,
respectively.

Using the charm veto events, we perform a cross-check of our
analysis.  We determine $B \to J/\psi K^{(*)}$ branching fractions
by following the
same analysis procedure with $3.07$ GeV/$c^2 < \mpp < 3.11$ GeV/$c^2$
and using ${\mathcal B}(J/\psi \to \pp) = ( 2.12 \pm 0.10 \times 10^{-3})$
~\cite{PDG}. The obtained branching fractions
are ${\mathcal B}(\bp \to J/\psi K^+) = ( 1.17^{+0.12}_{-0.13} \pm 0.16)
\times 10^{-3}$,
${\mathcal B}(B^0 \to J/\psi K^0)
= ( 1.16^{+0.24}_{-0.24} \pm 0.16)
\times 10^{-3}$,
${\mathcal B}(\bp \to J/\psi \kst) = ( 1.08^{+0.51}_{-0.42} \pm 0.17)
\times 10^{-3}$, and
${\mathcal B}(\bz \to J/\psi \ksz) = ( 1.40^{+0.27}_{-0.26} \pm 0.21)
\times 10^{-3}$,
which are in agreement with the world average values~\cite{PDG}.

The present results, shown in Table~\ref{tab:tot},
offer valuable information for understanding the mechanism of charmless
baryonic $B$ decay. In particular, the threshold peaking
behavior is
now firmly established. 
The $\bp \to \ppk$ data can be used to constrain the production of narrow
glueball states that decay to $\pp$~\cite{glueball}. With the current
statistics, we can not set a very stringent bound.
%
It should be noted that the observed $\bp\to\pppi$ rate is less
than $\bp\to\ppk$, which is consistent with what is observed in
$B\to K\pi$, $\pi\pi$ modes. The $\bz\to p\bar pK^0$ rate is
considerably lower than that of the $\bp\to\ppk$ mode
which should be contrasted with $B^{0,+}\to \pi^0 K^{0,+}$ modes
and also $B^{0,+}\to J/\psi K^{0,+}$ modes . 
This indicates that the intermediate $p\bar p$ system is nontrivial.
These modes are of interest for direct $CP$ violation searches. For
the $B^\mp \to p\bar pK^\mp$ and $p\bar p\pi^\mp$ modes that have
larger statistics, we define the charge asymmetry  
as $(N_{B^-} - N_{B^+})/ (N_{B^-} + N_{B^+})$ and find the values are 
$-0.05 \pm 0.11 \pm 0.01$ and  
$-0.16\pm0.22 \pm 0.01$ 
respectively.  
The systematic error is determined by checking the null asymmetry with 
a $B \to D (\to K \pi) \pi$ control sample. 
The measured asymmetries are consistent with zero for their large statistical
uncertainties.

We wish to thank the KEKB accelerator group for the excellent
operation of the KEKB accelerator.
We acknowledge support from the Ministry of Education,
Culture, Sports, Science, and Technology of Japan
and the Japan Society for the Promotion of Science;
the Australian Research Council
and the Australian Department of Education, Science and Training;
the National Science Foundation of China under contract No.~10175071;
the Department of Science and Technology of India;
the BK21 program of the Ministry of Education of Korea
and the CHEP SRC program of the Korea Science and Engineering Foundation;
the Polish State Committee for Scientific Research
under contract No.~2P03B 01324;
the Ministry of Science and Technology of the Russian Federation;
the Ministry of Education, Science and Sport of the Republic of Slovenia;
the National Science Council and the Ministry of Education of Taiwan;
and the U.S.\ Department of Energy.


\begin{thebibliography}{99}
\bibitem{conjugate} {Throughout this report,
inclusion of charge conjugate mode is always implied unless
otherwise stated.}
\end{thebibliography}
\end{document}